\documentclass[longauth]{aa}
\usepackage{color}
\usepackage{graphicx}
\usepackage{txfonts}
\usepackage[colorlinks, citecolor=blue]{hyperref}
\usepackage{mathtools}
\usepackage{ulem}
\usepackage{natbib}
\bibpunct{(}{)}{;}{a}{}{,} 
\usepackage{float}

\newcommand{\sio}{Si\,{\sc iv}/O\,{\sc iv}]}
\newcommand{\civ}{C\,{\sc iv}}
\newcommand{\ciii}{C\,{\sc iii}]}
\newcommand{\feii}{Fe\,{\sc ii}}
\newcommand{\mgii}{Mg\,{\sc ii}}
\newcommand{\neiii}{[Ne\,{\sc iii}]}
\newcommand{\nev}{[Ne\,{\sc v}]}
\newcommand{\oii}{[O\,{\sc ii}]}

\begin{document}

\title{Gaia GraL: {\it Gaia} DR2 Gravitational Lens Systems. V. \\Doubly-imaged QSOs discovered from entropy and wavelets\thanks{Some of the data presented herein were obtained at the W. M. Keck Observatory, which is operated as a scientific partnership among the California Institute of Technology, the University of California and the National Aeronautics and Space Administration. The Observatory was made possible by the generous financial support of the W. M. Keck Foundation. Partially based on observations collected at the European Southern
Observatory, Chile (104.A-0575).}}

   \author{A. Krone-Martins\inst{1,2},
           M.J. Graham\inst{3},
   		   D. Stern\inst{4},
           S.G. Djorgovski\inst{3},
           L. Delchambre\inst{5},
           C. Ducourant\inst{6},
           R. Teixeira\inst{7},\\
           A.J. Drake\inst{3},
           S. Scarano Jr.\inst{8},
           J. Surdej\inst{5},
           L. Galluccio\inst{9},
           P. Jalan\inst{10,11},
           O. Wertz\inst{12},
           J. Kl\"{u}ter\inst{13},  
           F. Mignard\inst{9},\\
           C. Spindola-Duarte\inst{7},
           D. Dobie\inst{14,15},
           E. Slezak\inst{9},
           D. Sluse\inst{5},
           T. Murphy\inst{14},
           C. Boehm\inst{14},\\
           A. M. Nierenberg\inst{4},
           U. Bastian\inst{13},
           J. Wambsganss\inst{13,16},
           J.-F. LeCampion\inst{6} 
        }

  \institute{
  	     CENTRA, Faculdade de Ci\^encias, Universidade de Lisboa, Ed. C8, Campo Grande, 1749-016 Lisboa, Portugal\\
          \email{algol@sim.ul.pt}
        \and
         Donald Bren School of Information and Computer Sciences, University of California, Irvine, 
         Irvine CA 92697, USA
        \and
         California Institute of Technology, 1200 E. California Blvd, Pasadena, CA 91125, USA
        \and
         Jet Propulsion Laboratory, California Institute of Technology, 4800 Oak Grove Drive, Pasadena, CA 91109, USA
        \and
         Institut d'Astrophysique et de G\'{e}ophysique, Universit\'{e} de Li\`{e}ge, 19c, All\'{e}e du 6 Ao\^{u}t, B-4000 Li\`{e}ge, Belgium
        \and          
         Laboratoire d'Astrophysique de Bordeaux, Univ. Bordeaux, CNRS, B18N, all{\'e}e Geoffroy Saint-Hilaire, 33615 Pessac, France
        \and
         Instituto de Astronomia, Geof\'isica e Ci\^encias Atmosf\'ericas, Universidade de S\~{a}o Paulo, Rua do Mat\~{a}o, 1226, Cidade Universit\'aria, 05508-900 S\~{a}o Paulo, SP, Brazil
        \and
         Departamento de F\'{i}sica – CCET, Universidade Federal de Sergipe, Rod. Marechal Rondon s/n, 49.100-000, Jardim Rosa Elze, S\~{a}o Crist\'{o}v\~{a}o, SE, Brazil
        \and
         Universit\'{e} C\^{o}te d'Azur, Observatoire de la C\^{o}te d'Azur, CNRS, Laboratoire Lagrange, Boulevard de l'Observatoire, CS 34229, 06304 Nice, France
        \and
         Aryabhatta Research Institute of Observational Sciences (ARIES), Manora Peak, Nainital 263002, India
        \and
         Department of Physics and Astrophysics, University of Delhi, Delhi 110007, India
        \and
         Argelander-Institut f\"{ u}r Astronomie, Universit\"{ a}t Bonn, Auf dem H\"{ u}gel 71, 53121 Bonn, Germany
        \and
         Zentrum f\"{u}r Astronomie der Universit\"{a}t Heidelberg, Astronomisches Rechen-Institut, M\"{o}nchhofstr. 12-14, 69120 Heidelberg, \\Germany
        \and
         School of Physics, The University of Sydney, NSW 2006, Australia
        \and
         ATNF, CSIRO Astronomy and Space Science, PO Box 76, Epping, New South Wales 1710, Australia
        \and
         International Space Science Institute (ISSI), Hallerstra\ss e 6, 3012 Bern, Switzerland     
        }

   \date{Received December 20, 2019; accepted ???, ???}

%%%%% ABSTRACT %%%%%
  \abstract
   {The discovery of multiply-imaged gravitationally lensed QSOs is fundamental to many astronomical and cosmological studies. However, these objects are rare and challenging to discover due to requirements of high-angular resolution astrometric, multiwavelength photometric and spectroscopic data. This has limited the number of known systems to a few hundred objects.}
  % aims heading (mandatory)
   {We aim to reduce the constraints on angular resolution and discover multiply-imaged QSO candidates by using new candidate selection principles based on unresolved photometric time-series and ground-based images from public surveys.}
  % methods heading (mandatory)
    {We selected candidates for multiply-imaged QSOs based on low levels of entropy computed from Catalina unresolved photometric time-series or Euclidean similarity to known lenses in a space defined by the wavelet power spectra of Pan-STARSS DR2 or DECaLS DR7 images, combined with multiple {\it Gaia} DR2 sources or large astrometric errors and supervised and unsupervised learning methods. We then confirmed spectroscopically some candidates with the Palomar Hale, Keck-I, and ESO/NTT telescopes.}
  % results heading (mandatory)
     {We report the discovery and confirmation of seven doubly-imaged QSOs and one likely double quasar. This demonstrates the potential of combining space-astrometry, even if unresolved, with low spatial-resolution photometric time-series and/or low-spatial resolution multi-band imaging to discover multiply-imaged lensed QSOs.}
   {}

\titlerunning{{\it Gaia} GraL V -- Doubly-imaged QSOs discovered from entropy and wavelets}
\authorrunning{A. Krone-Martins et al.}
\maketitle

%%%%% INTRODUCTION %%%%%
\section{Introduction}
Strong gravitational lenses (GLs) originate from a simple light deflection phenomenon: the bending of the light by massive intervening objects, which upon favorable alignment, can lead to the formation of multiple images of a background source. The quantitative descriptions of this phenomenon \citep{1936Sci....84..506E, 1937PhRv...51..290Z} were only possible after General Relativity. It then took many decades before it was possible to obtain an observational proof that this phenomenon indeed happens in the Universe \citep{WCW_firstGL_1979}. Even nowadays, only a few tens of quadruply-imaged and barely two hundred doubly-imaged QSO GLs are known \citep[e.g.][]{Ducourant_GaiaGraL_paper_II_2018, 2019MNRAS.483.4242L}. The majority of these systems were detected using variations of the following procedure: first candidates are pre-selected from catalogs, either from spatially resolved sources close together having similar colors, or by some perturbation of its measurements (e.g. large astrometric and photometric errors), or by certain other color criteria; then, images are inspected visually or algorithmically and selected based on geometric patterns; finally, the most promising candidates are spectroscopically observed.

With a few exceptions \citep[e.g.][]{2005ApJ...624L..21B, 2010ApJS..188..280S, 2012A&A...540A..36C, 2018MNRAS.476..663K}, this is still how most GLs are discovered today. More recently, the aforementioned described procedure has been optimized so it can be applied to large datasets produced by astronomical surveys. This is being done for instance by optimizing spatial pattern queries in large databases  \citep[e.g.][]{DBLP:conf/ssdbm/PortoROKVS18}, and by exploiting the predictive power of machine learning methodologies \citep[e.g.][]{2015MNRAS.448.1446A, 2018MNRAS.473.3895L, Delchambre_GaiaGraL_paper_III_2018, 2019arXiv190601638K, 2019arXiv191104320C}, leading to the discovery of new GLs \citep[e.g.][]{2017MNRAS.465.4325O, 2018A&A...616L..11K, 2018arXiv181002624W}. However, these selection principles require resolved candidates, whereas a large number of lenses will not be spatially resolved by modern ground-based surveys \citep[e.g.][]{2010MNRAS.405.2579O,2016Finet}. On the other hand, Astronomy is already in an era of time-resolved imaging all-sky surveys, some of them being multiwavelength. This realm was opened by the Catalina Real-Time Transient Survey \citep[][CRTS]{2009ApJ...696..870D}, and it is now being enlarged with the Hyper Suprime-Cam Subaru Strategic Program \citep[][HSC]{2018PASJ...70S...8A}, the Zwicky Transient Facility \citep[][ZTF]{2019arXiv190201945G} and soon by the Large Synoptic Survey Telescope \citep[][LSST]{2008arXiv0805.2366I}. Although there have been some proposals to leverage the enormous power of time-resolved photometric and unresolved imaging to assist the discovery of new GLs \citep{2006ApJ...637L..73K, 2019arXiv191001140C}, there is still a lack of exploration and methodological development to profit from these data. The aim of this Letter is to contribute to this issue by describing two simple selection principles to benefit from the power of these surveys. These  principles hold even when the time-resolved surveys cannot spatially resolve the candidates. 

In Sect.~\ref{sec:data} we describe the data adopted in this work. In Sect. \ref{sec:principles} we describe two new candidate selection principles. In Sect.~\ref{sec:observations} we describe the observations and the confirmation of new multiply-imaged systems. Finally, we conclude in Sect.~\ref{sec:conclusions}. Throughout this work we adopt $\Omega_m = 0.315$, $\Omega_\Lambda = 0.685$ and $H_0 = 67.4$ km\, sec$^{-1}$\, Mpc$^{-1}$ \citep{2018arXiv180706209P}.

\section{Data}
\label{sec:data}
In this work, we first created a list of confirmed or candidate QSOs by merging the Milliquas v.5.7b \citep{2015PASA...32...10F} and the {\it WISE} AGN R90 catalogs \citep{2018ApJS..234...23A}. This initial list of QSOs was then adopted to extract lightcurves and images.

Time-resolved photometry was extracted from CRTS. We selected $\sim1.8\times10^5$ lightcurves from CRTS detections that had known QSOs counterparts in our initial list, excluding sources classed as blends, sources with a nearby bright star and known blazars. Imaging was extracted from the Pan-STARRS1 \citep[][PS1DR2]{2016arXiv161205560C} Data Release 2 survey\footnote{http://ps1images.stsci.edu/cgi-bin/fitscut.cgi} and from the Dark Energy Camera Legacy Survey Data Release 7 \citep[][DECaLS]{2019AJ....157..168D}\footnote{http://legacysurvey.org/viewer/jpeg-cutout}. Color images in the $g$, $r$ and $z$ bands were obtained as compressed ISO/IEC-ITU JPEG format \citep[e.g.][]{10.1117/1.JEI.27.4.040901}, $\sim10$\arcsec\ wide, resulting in $\sim 4.7\times10^{6}$ images from PS1DR2 and  $\sim 2.4\times10^{6}$ images from DECaLS.

As CRTS has FWHMs of $\sim3"-6"$ with pixel sizes of $\sim1"-3"$ and PS1DR2 and DECaLS are seeing limited surveys, we also adopted data from the {\it Gaia} space mission \citep{2016A&A...595A...1G} in this first study. We extracted the {\it Gaia} Data Release 2 data \citep[][{\it Gaia} DR2]{2018Gaia} from the list of QSOs using the {\it Gaia} archive  \citep{2017A&C....21...22S} via an ADQL cross-match. In the specific case of the CRTS, we only selected lightcurves from sources containing at least two independent {\it Gaia}  counterparts within a radius of 2\arcsec\ from the CRTS position (roughly, the CRTS resolution), as our objective is to test scenarios with spatially unresolved time-domain data.

\section{The new principles and candidate selection}
\label{sec:principles}
\subsection{Lightcurve entropy selection principle}
\label{sec:entropyprinciple}
If the time-resolved survey, in our case CRTS, had enough spatial resolution to derive two independent lightcurves, the detection of a possible multiply imaged quasar would be trivial. If the sampling is sufficient, a simple comparison of the lightcurves using some distance metric would reveal that they are likely the same lightcurve except for minor effects such as microlensing \citep[e.g.,][]{Wambsganss_Q2237_1991, Chae_H1413_2001, Akhunov_Wertz_2017}. Although microlensing can harm inferences of $H_0$, it cannot prevent the identification of the two time-series as realizations of the same underlying time-series, as expected from a gravitational lens. However, time-resolved all-sky surveys do not attain sufficient spatial resolution to resolve most lenses \citep[e.g.][]{2010MNRAS.405.2579O,2016Finet}. 

The resulting unresolved lightcurve of a multiply-imaged QSO is therefore the addition of multiple copies of a single stochastic time-series with a time-lag. Thus, this lightcurve will be less stochastic than the individual lightcurves, as its autocorrelation is increased, and thus it will present lower entropy. Accordingly, as there is no reason to expect that physical mechanisms ruling the stochastic behavior of QSO emission are different between the population of lensed QSOs and the population of non-lensed QSOs, the distribution of some measurement of entropy of the lightcurve calculated from the population of unresolved multiply-imaged QSOs should be more concentrated towards lower values than the distribution of non-lensed QSOs.

The construction of a simple method to select candidates based on this principle is then straightforward: some entropy measurement is calculated from a set of QSO lightcurves and the lower entropy sources are selected as candidates. In the first year of the {\it Gaia} Gravitational Lens, or GraL, program \citep{2018A&A...616L..11K}, we adopted the multivariate multiscale sample entropy \citep{PhysRevE.84.061918}, as it can take into account multiple wavelengths as expected from ZTF and LSST. We used the \texttt{MSMVSampEn} R implementation\footnote{https://github.com/areshenk/MSMVSampEn}. This measurement is sensitive to the time series sampling, thus we only considered well-sampled QSOs and resampled the lightcurves to regular intervals. We also only selected sources with more than two {\it Gaia} DR2 detections in this first test. Then, we trained a Support Vector Machine (SVM) with a radial basis function kernel \citep{Boser1992,Cortes95support-vectornetworks} on the CRTS entropy and {\it Gaia} DR2 angular separation space using known lenses. Finally, we applied this method to select new candidates, which were visually inspected before spectroscopic follow-up. 

\subsection{Image wavelet powerspectra selection principle}
If multiwavelength images could be obtained with infinite spatial resolution, multiply-imaged QSOs would be easily resolved and detected. Yet, even if these objects are not fully resolved, barely resolved images should contain signatures hinting to their nature. For instance, some extension of the point-spread function could indicate more than one QSO image, mixed with spatial color signature, that could hint to a possible lens. These signatures could be revealed, then, by multiscale analysis.

In such a multiscale context, wavelet transforms \citep[e.g.][]{Haar1910, GOUPILLAUD198485} have been long adopted in astronomical signal processing and analysis \citep[e.g.][]{1990A&A...227..301S, 1994A&A...288..342S}, usually for their properties to denoise, enhance and separate n-dimensional signals. Here, however, we adopt wavelets in another scenario: information retrieval. We use the wavelet representation of an image as a proxy for its information content, and thus we index QSO images by their wavelet space representation and cast the lens search problem as a reverse image search, or content-based image retrieval \citep[e.g.][]{Long2003}. Thus, even if the source catalog produced from the survey images cannot resolve individual sources, we can use similarity searches in the wavelet space or from features derived from the wavelet representation to detect similar spatial and color (in multiwavelength data) structures between different sets of images. In the aim of effectiveness, one should require these structures to be translation and rotation invariant. Thus, here we used the total power contained at each wavelet scale.

A simple method can then be constructed from this principle: (i) Wavelet decompositions of an image set are computed (ii) For each object, the total power at each wavelet scale is calculated for each passband and each adjacent color for a multiwavelength survey (e.g. $g$, $r$, $i$, $z$, and $y$ passbands, and $g-r$, $r-i$, $i-z$, and $z-y$ colors in PS1DR2), resulting in the equivalent of a power spectrum. (iii) For each object, a vector is created concatenating and normalizing the powerspectra per band and color. (iv) Finally, these vectors can be clustered according to Euclidian distances, resulting in a tree structure. It is then trivial to perform similarity searches by walking through the nearest branches of the tree.

In the context of the GraL lens search, we built such clustered maps of vectors from PS1DR2 and DECaLS images, extracted around previously known QSOs or candidates, regardless of their detection by the survey catalogs. For the purposes of the first year of our lens search, we performed the 2D wavelet analysis using the \citet{Mallat:1989:TMS:67253.67254} pyramidal algorithm and a Daubechies Least-Asymmetric wavelet with four vanishing moments \citep[][]{Daubechies:1993:OBC:154993.155007} as implemented by the \texttt{wavethresh} R package \citep[][]{Nason:2008:WMS:1481436}. For this first test we selected the objects closer in the L2-norm sense to a set of lenses selected for their clear morphology\footnote{The lenses were: PG1115+080 \citep{1980Natur.285..641W}, 
SDSS0924+0219 \citep{2003AJ....126..666I}, 
SDSS0246-0825  \citep{2005AJ....130.1967I}, 
SDSSJ105440.83+273306.4 \citep{2012AJ....143..119I}, 
SDSSJ1537+3014 \citep{2016More}, 
WISE025942.9-163543 \citep{2017AJ....153..219S}, 
HSCJ115252+004733 \citep{2017MNRAS.465.2411M}, 
WGD0245-0556 \citep{2018MNRAS.479.4345A}, 
J0941+0518 \citep{2018MNRAS.477L..70W}, J0011-0845 \& J0123-0455 \citep{2018MNRAS.479.5060L} and GraL~J175443398+214054818 \citep{Delchambre_GaiaGraL_paper_III_2018}.}. From those, we selected objects with a good astrometric behavior if multiple sources were detected in {\it Gaia} DR2, or bad astrometric behavior. The reason for is that badly behaved astrometric solutions can indicate that the {\it Gaia} DR2 solution was derived from data of more than one point source. Finally, the selected objects had their archival images visually inspected before scheduling spectroscopic observations. 

%%%%%%%%%%%%%%%%% TABLE ----
   \begin{table*}
      \caption[]{Observations of the confirmed GraL doubly-imaged QSOs and possibly double QSOs selected from entropy and wavelet powerspectra. Conservative redshift uncertainties for the Keck estimates are at the $\sigma z\sim0.002$ level.}
         \label{DoublyImagedQSOs}
         \small
         \centering
         \begin{tabular}{lcccccl}
            \hline
            \noalign{\smallskip}
            Name & Telescope & UT Epoch & P.A. & Exp. Time (s) & $z$ & Notes\\
            \noalign{\smallskip}
            \hline
            \noalign{\smallskip}
GraL~J024612.2$-$184505.1 & Keck    & 2019 Feb 06 &   0$^{\circ}$ &          $1\times300$ & 1.873 & \\ % ``Eridanus Tortoises''\\ %close separation
                          & ATCA    & 2019 Dec 09 &    &           &   & detection at 9~GHz\\

GraL~J034611.0+215444.6   & Palomar & 2018 Aug 20 &  27$^{\circ}$ & $2\times300$ & 2.355 & \\ %``Taurus Courage'' \\

GraL~J081830.5+060137.9   & Keck    & 2019 Feb 06 &  60$^{\circ}$ & $3\times300$ & 2.352 & absorption at $z=1.007, 1.507$ \\ % ``Hydra's Tongue''\\  % close separation 
%   &     &  &   & &  & Multiple absorption systems\\
% &  &     &  &   & &  & absorption at $=1.007$ \& $z=1.507$ \\  % close separation 
GraL~J090710.5+000321.7   & Keck    & 2019 Feb 06 & 160$^{\circ}$ & $3\times600$ & 1.299 & absorption at $z=0.682, 0.771$\\
% % ``Hydra's Heart''\\%close separation \\
%   &     &  &   & &  & absorption at $z=0.682$ \& $z=0.771$\\  % close separation 
GraL~J125955.5+124152.6   & Keck    & 2019 Jun 01 & 220$^{\circ}$ & $2\times600$ & 2.196 & stronger self-absorption in one image \\
% \\ % ``Bette Davis'' or ``Virgo's Pearls''\\
GraL~J155656.1$-$135210.1 & Keck    & 2019 Feb 06 &  30$^{\circ}$ & $2\times300$ & 1.423 & absorption at $z=0.393$\\ % ``Libra's Weight'' \\
                          & NTT    & 2019 Apr 09 & $20^{\circ}$ & $1\times1320$ & $1.43\pm0.01$ &  \\

GraL~J220015.6+144859.5   & Keck    & 2019 Jun 01 & 349$^{\circ}$ & $2\times600$ & 1.115 & absorption at $z=0.762$ \\ %``Pegasus Jewels'' \\
GraL~J234330.6+043557.9   & Palomar & 2018 Aug 20 & 110$^{\circ}$ & $2\times600$ & 1.604 & absorption at $z=0.855$ \\ %``Poseidon's Blue Balls'' \\ %close separation
                          & Keck    & 2018 Sep 15 & 117$^{\circ}$ & $3\times300$ & 1.604 & \\ %``Frank Sinatra''
     		          & Keck    & 2019 Jan 12 & 117$^{\circ}$ & $3\times300$ & 1.604 &  \\
            \hline
         \end{tabular}
   \end{table*}
%%%%%%%%%%%%%%%%% TABLE ----

%\section{Observations and the new lenses}
\section{Observations and validation of the new lenses}
\label{sec:observations}
%--------------------------------------
The observations performed during the first year of the GraL lens search targetted some candidates identified from methods based on the new entropy and wavelet power spectra selection principles. Here we report on such confirmed doubly-imaged QSOs. Stern et al. (in prep.; GraL Paper VI) provides a fuller discussion of all GraL spectroscopic observations to date, including targets from multiple selection methods and reporting  confirmed lenses, quadruply-imaged QSOs, ambiguous and failed candidates.% where one or both of the targets were confirmed as Galactic stars.

%The observations performed during the first year of the GraL lens search targetted some candidates that were independently selected using the search methods described in the previous section. Here we report confirmed lensed quasars identified from methods based on the new entropy and wavelet power spectra selection principles. Stern et al. (in prep.; GraL Paper VI) provides a fuller discussion of all GraL spectroscopic observations to date, including targets from multiple selection methods and reporting successfully confirmed lenses, ambiguous and failed candidates.% where one or both of the targets were confirmed as Galactic stars.

%Observations of confirmed entropy and wavelet power spectra selected candidates were carried out between 2018 August and 2019 June using three instruments: 

\begin{figure*}[!ht]
\begin{center}
\includegraphics[trim={0cm 5cm 0cm 5cm},clip,width=0.67\textwidth]{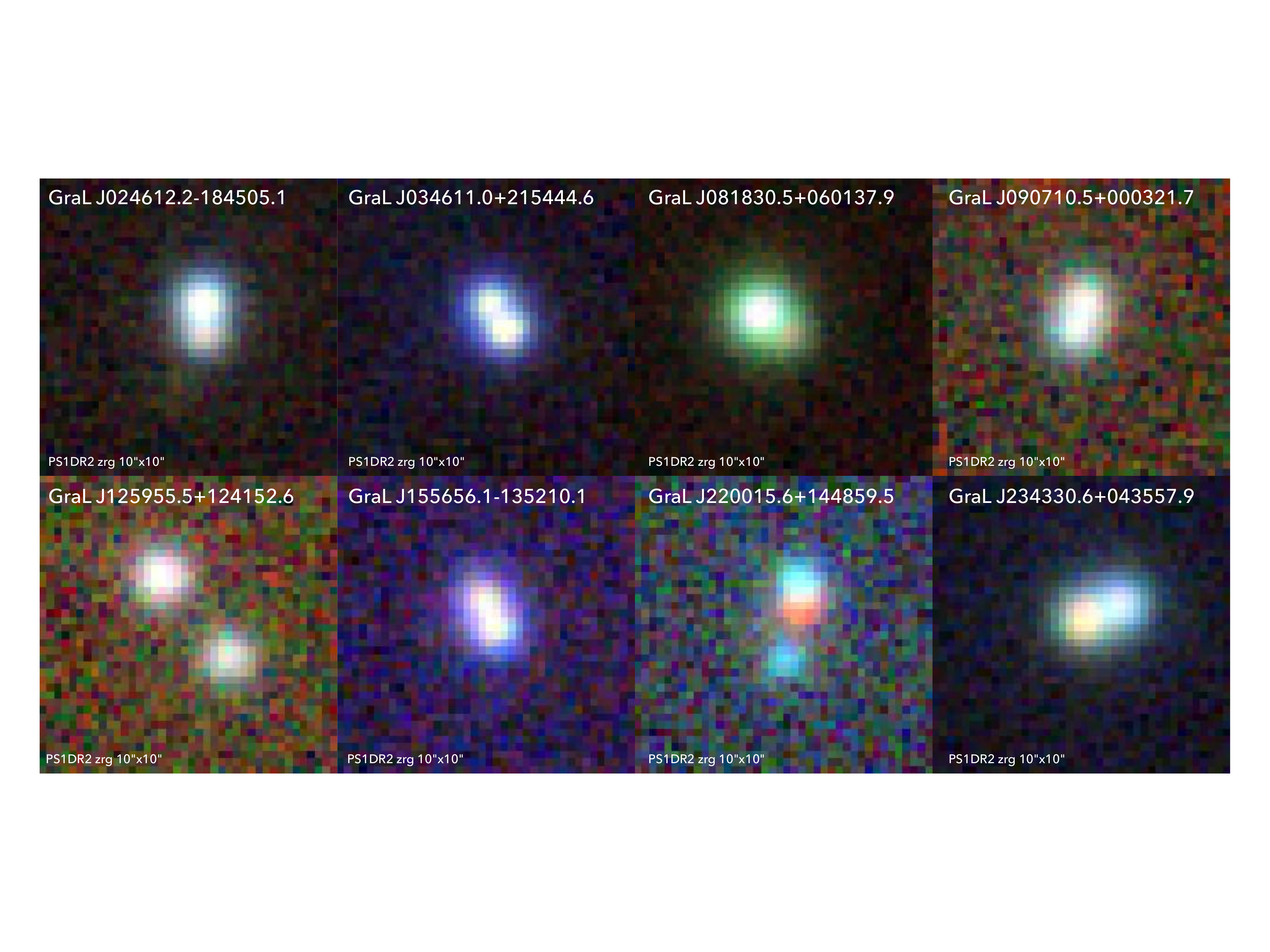} 
\caption{PS1DR2 images, 10\arcsec\ on a side, of the multiply-imaged QSOs and the likely double QSO (GraL~J125955.5+124152.6).  North is up, and east is to the left. The false RGB colors were created from the $z'$, $r'$, and $g'$ filters, respectively.
\label{fig:Stamps}}
\end{center}
\end{figure*}

Observations of the selected candidates were carried out between 2018 August and 2019 June using three instruments: the dual-beam Double Spectrograph \citep[][DBSP]{1982PASP...94..586O} on the 5-m Palomar Hale telescope, the dual-beam Low-Resolution Imaging Spectrometer \citep[][LRIS]{Oke_LRIS_1995} on the 10-m Keck~I telescope, and the European Southern Observatory (ESO) Faint Object Spectrograph and Camera \citep[][EFOSC2]{1984Msngr..38....9B, 2008Msngr.132...18S} on the 3.6-m New Technology Telescope (NTT). Over this period we selected a total of 44 entropy and 5 wavelet power spectra candidates for observation. The low number of the latter was because wavelet candidates only started to be selected in mid-2019. The 23 Palomar candidates were selected based on a direct entropy-level cut, while Keck and NTT candidates adopted the simple method described in Sect.~\ref{sec:entropyprinciple}. Table~\ref{DoublyImagedQSOs} presents details on the observations: telescopes used, observation epochs, slit position angles (P.A.'s), exposure times, derived lensed quasar redshifts and additional notes. Figure~\ref{fig:Stamps} presents PanSTARRS false-color optical images of the sources discussed in this work. Here we briefly describe the observations and data reductions, and the following subsection provides notes on the individual confirmed lenses. 

{\bf Palomar Hale.} Two eventually confirmed candidates were observed with Palomar/DBSP on UT 2018 August 20, a photometric night with 1\arcsec\ seeing.  We used the 1\arcsec\ width slit, the 5600\AA\ dichroic, the 600~$\ell$~mm$^{-1}$ blue grism ($\lambda_{\rm blaze}
= 4000$~\AA), and the 400~$\ell$~mm$^{-1}$ red grating ($\lambda_{\rm blaze} = 8500$~\AA).  This instrument configuration covers the full optical window at moderate resolving power, $R \equiv \lambda / \Delta \lambda \approx 1250$. Standard stars BD+33~2642 and Feige
110 were observed the same night for flux calibration.

{\bf Keck-I.} Additional candidates were observed with Keck/LRIS on the nights of UT 2018 September 15, UT 2019 January 12, UT 2019 February 6, and UT 2019 June 1. Conditions on all nights were clear with $\sim 1\arcsec$ seeing.  We used the 1\farcs0 and 1\farcs5 width slits, the 600 $\ell$~mm$^{-1}$ blue grism ($\lambda_{\rm blaze} = 4000$~\AA), and the 400 $\ell$~mm$^{-1}$ red grating ($\lambda_{\rm blaze} = 8500$~\AA). The first three runs used the 5600~\AA\, dichroic, while the final run used the 6800~\AA\, dichroic. These instrument configurations cover the full optical window at moderate resolving power, $R \approx 1000$ for the wider slit and $R \approx 1500$ for the narrower slit (for objects filling the slit). Over the course of these nights, standard stars from \citet{1990ApJ...358..344M} were observed for flux calibrating each night. %of observations.
%\citet{1990ApJ...358..344M} standard stars were observed for flux calibrating each night.

{\bf ESO/NTT.} Another candidate validation session was performed with ESO/NTT/EFOSC2 on the nights of UT 2019 April 7-9. Conditions were clear with $\sim 1\arcsec$ seeing. We observed 9 candidates using the 5\farcs0 and 1\farcs5 width slits, the Grism1 covering 3185-10940~\AA~($\lambda_{\rm blaze} = 4500$~\AA) and the GG375 order-blocking filter. The LTT3864 and LTT7379 spectrophotometric standards  \citep{1994PASP..106..566H} were observed in the first two nights.
%STANDARD STARS LINK : https://www.eso.org/sci/observing/tools/standards/spectra/stanlis.html

%{\bf ESO/NTT.} Another session of validation of candidates was performed at ESO/NTT/EFOSC2 on the nights of 2019 April 7-9. Conditions on all nights were clear with $\sim 1\arcsec$ seeing. For the ESO/NTT/EFOSC2 observations, we used the 5\farcs0 and 1\farcs5 width slits together with Grism1 covering 3185-10940~\AA ($\lambda_{\rm blaze} = 4500$~\AA) and filter GG375. 

All observations were processed using standard techniques within \texttt{IRAF}, and flux-calibrated using observations of the aforementioned spectrophotometric standards.% observed on the same nights.

\subsection{Individual Objects}

{\bf GraL~J024612.2$-$184505.1 ---} This is a close separation ($\sim$1\farcs001), possibly lensed quasar. {\it Gaia} DR2 indicates two sources, the brighter at $G\sim18.4$ mag, and  fainter/brighter flux ratio of $0.440\pm0.006$. PS1DR2 and DECaLS images cleary show spatial extent (e.g. see Figure \ref{DoublyImagedQSOs}), but the PS1 catalog contains a single source. The confirming Keck spectrum shows a clear spatial extent but fails to separate the components. Considering that neither absorption nor evidence for the lensing galaxy was detected in the current spectroscopic data, binarity cannot be excluded yet. No radio or X-ray counterparts are reported in the NASA Extragalactic Database (NED). This source was followed up on the UT 2019 December 9 with the Australia Telescope Compact Array \citep[][ATCA]{2011MNRAS.416..832W}, and the preliminary data reduction shows a detection at 9 GHz. 

{\bf GraL~J034611.0+215444.6 ---} This newly identified lensed quasar consists of a close, sub-arcsecond ($\sim$0\farcs988), separation pair of blue sources. {\it Gaia} DR2 indicates two sources, the brighter at $G\sim18.8$ mag, and flux ratio $0.812\pm0.007$. PS1DR2 images show spatial extent, but the catalog contains a single source. The confirming Palomar spectrum shows a spatial extent but fails to separate the components, except at the reddest wavelengths. We observe that blueward of the \civ\ emission line, there is a deep, narrow absorption line almost reaching zero flux. This is only possible if the two components show a similar absorption, confirming that this system is most likely a doubly-imaged quasar. No radio or X-ray counterparts are reported in NED.

{\bf GraL~J081830.5+060137.9 ---} The confirming Keck spectrum shows a clear spatial extent, but fails to separate the two components. {\it Gaia} DR2 indicates two close sources ($\sim$1\farcs147), the brighter at $G\sim17.8$ mag, and flux ratio $0.160\pm0.001$. PS1DR2 and DECaLS images show spatial extent, but the PS1 catalog contains a single source. The quasar is a broad absorption line (BAL) quasar, with multiple absorption components observed out to $\sim 5000\, {\rm km}\, {\rm s}^{-1}$. Multiple foreground \mgii\ absorption systems are also seen, including absorption from both \mgii\ and \feii\ at $z = 1.507$ and \feii\ absorption at $z = 1.007$ (with \mgii\ from the lower redshift system lost to the dichroic). This source was reported by the SDSS-III BOSS quasar lens survey \citep{2016More}, though they consider it to be a quasar pair rather than a confirmed lensed quasar based on the non-detection of a lensing galaxy in their moderate-depth optical follow-up imaging (420~s $i$-band image from the SOAR 4-m telescope). They note that if this quasar is lensed, then the non-detection would imply a high-redshift lensing galaxy, at $z > 1$, which would be consistent with the foreground absorption systems detected here if one of those systems is lensing the background quasar. This source is also independently identified as a lens by \citet{2019arXiv191204336H}. They reported the presence of the absorption system at $z=1.007$. Assuming that the lens is at the absorber redshift $z=1.507\, (1.007)$ and assuming $\theta_E$ to be half the angular image separation of 1\farcs15, we derive a mass enclosed within $\theta_E$ of $\sim1.3\times 10^{12}\, (6.8\times10^{11})\, M_\odot$ for the deflector and a proper comoving length larger than $\gtrapprox25 (19)$ kpc for the absorber cloud. The $\Delta z = 0.01$ discrepancy between our redshift and \citet{2016More} is not surprising given the challenge of determining the systemic redshift of BAL quasars and the different measurement methods. No radio or X-ray counterparts are reported by NED. 

{\bf GraL~J090710.5+000321.7 ---} This is a newly identified close separation ($\sim$0\farcs983) lensed quasar. {\it Gaia} DR2 indicates two sources, the brighter at $G\sim20.2$ mag, and flux ratio $0.84\pm0.02$. PS1DR2 and DECaLS images show spatial extent, but the PS1 catalog contains a single source. The confirming Keck spectrum shows a clear spatial extent but fails to separate the components. Absorption lines from \mgii\ and \feii\ are clearly seen from two foreground systems, at $z = 0.682$ and $z = 0.771$. Similar to GraL~J034611.0+215444.6, here we also observe narrow absorption lines (cf. CIV) almost reaching zero flux. Assuming that the lens is located at the intervening absorber redshift $z=0.682\, (0.771)$ and given the angular separation of $\sim$0\farcs98, we derive a mass of $\sim 4.7 \times 10^{11}(5.9\times10^{11})\, M_\odot$ for the deflector and a proper comoving length $\gtrapprox 12\, (13)$ kpc for the absorber cloud. No radio or X-ray counterparts are reported in NED.

{\bf GraL~J125955.5+124152.6 ---}  This system, the widest separation system in this sample ($\sim$3\farcs513), is associated with the bright {\it ROSAT} X-ray source 1WGA~J1259.9+1241, and is listed in the {\it Chandra} Source Catalog \citep{2010ApJS..189...37E}. It was reported as a binary quasar in \citet{2006AJ....131....1H} and as a double quasar in \citet{2009ApJ...693.1554F}. {\it Gaia} DR2 indicates two sources, the brighter at $G\sim19.7$ mag, and flux ratio $0.896\pm0.006$. PS1DR2 and DECaLS images cleary show two sources, and the PS1 catalog also contains two sources. The Keck spectrum shows strong self-absorption in one of the two components. We note that ambiguity in whether this source is truly a lensed quasar, versus a double, or binary quasar. In particular, the entropy method will identify elevated entropy from double quasars as compared to single quasars, regardless of whether they are distinct sources or two lensed images of the same source with a temporal delay. Though the same ambiguity is true for other sources, the wider separation of this pair with no lensing galaxy apparent in imaging to date makes this the most likely non-lensed, double quasar, in this sample.

{\bf GraL~J155656.1$-$135210.1 ---} The confirming Keck spectrum shows clear spatial extent, but fails to separate the two components. Absorption lines are seen from \mgii\ and \feii\ at $z = 0.393$, potentially associated with the lensing galaxy. {\it Gaia} DR2 indicates two sources ($\sim$0\farcs959 separation), the brighter at $G\sim19.3$ mag, and flux ratio $0.609\pm0.004$. PS1DR2 images show two sources but the PS1 catalog contains a single source. The Keck conclusions are corroborated by the  reduction of NTT observations. The spectra are well separated in both observations. Their similarity suggests a doubly-imaged quasar at redshift $z=1.42$, but binarity cannot be completely excluded. Assuming that the lens is located at the absorber redshift $z=0.393$ and given the angular separation of $\sim$0\farcs9 between the lensed components, we derive a mass of $\sim1.8\times10^{11} M_\odot$ for the deflector and a proper comoving length for the absorber cloud $\gtrapprox7$ kpc. No radio or X-ray counterparts are reported in NED. 

{\bf GraL~J220015.6+144859.5 ---} The Keck spectrum of this wavelet-selected lensed quasar shows two distinct components, with one spectrum contaminated by a Galactic M-star, visible as the red source in Figure~\ref{DoublyImagedQSOs}. Absorption lines from \mgii\ and \feii\ are clearly seen in both lensed images at $z = 0.762$, potentially associated with the lensing galaxy. {\it Gaia} DR2 indicates one source at $G\sim20.5$ mag with large astrometric errors. PS1DR2 and DECaLS images, and the PS1 catalogue, contain two sources ($\sim$2\farcs5 separation). Assuming that the lens is located at the absorber redshift and given the angular separation of between the two lensed images, we derive a mass of $\sim4.9\times10^{12} M_\odot$ for the deflector and a proper comoving length for the absorber $\gtrapprox33$ kpc. No radio or X-ray counterparts are reported in NED.

{\bf GraL~J234330.6+043557.9 ---}  This newly confirmed lens consists of a close separation pair ($\sim$1\farcs231). {\it Gaia} DR2 indicates two sources, the brighter at $G\sim18.7$ mag, and flux ratio $0.778\pm0.009$. PS1DR2 and DECaLS images cleary show two sources, but the PS1 catalog contains a single source. The initial Palomar spectrum shows a spatially extended $z=1.604$ broad-lined quasar, separated into two components only at the reddest optical wavelengths and with one lensed component showing stronger absorption lines. The Keck follow-up observations clearly resolve the two images at all wavelengths, and identify multiple absorption lines at $z=0.855$ which appear much stronger in one of the lensed quasar components, which also is redder in color (see Appendix). % (Fig.\ref{fig:spectra2343blue}). 
It is assumed that this foreground absorption corresponds to the lensing galaxy, with one of the lensed quasar images intercepting a dense, dusty region of the lensing galaxy, such as a giant molecular cloud. Given the angular separation of $\sim$1\farcs23 between the two lensed components, we estimate a mass of $\sim8.6\times10^{11} M_\odot$ for the deflector and a proper comoving length for the covering absorber cloud $\lessapprox28$ kpc. No radio or X-ray counterparts are reported in NED. 

\section{Conclusions}
\label{sec:conclusions}

In this Letter we present two new selection principles for the discovery of multiply-imaged QSOs in the era of time-resolved, and large-scale, ground-based imaging astronomical surveys. 

The first selection principle is based on the stochastic behavior of QSO lightcurves, and that the lightcurve of a multiply-imaged QSO is the replication of a single stochastic lightcurve delayed in time. Thus if the images cannot be individually resolved, the entropy of the unresolved lightcurve will have lower values than the entropy of the general QSO population.

The second selection principle is based on the fact that even if the multiple images of lensed QSOs cannot be fully resolved, the unresolved or barely resolved image of the lensed system can present lensing signatures. Thus a multiscale analysis of the object image can reveal more power spread on different scales than the general QSO population, and moreover, it can be used as an image indexing scheme for a reverse image search. 

We applied these selection principles, at the moment implemented in exploratory methods, to select multiply-imaged QSO candidates from a list of known or candidate QSOs contained in the {\it Gaia} DR2, CRTS, PS1DR2 and DECaLS surveys. We observed these candidates using Palomar/DBSP, Keck/LRIS and ESO/NTT/EFOSC2, resulting in the discovery of seven new gravitationally lensed and a doubly-imaged QSO. One of these sources was also recently followed up in radio with ATCA, and larger radio, X-ray and optical IFU follow-up programs are currently in preparation for the ATCA, {\it XMM-Newton}, Gemini South/GMOS, and Keck/OSIRIS facilities. We are also developping further methodological work around these new selection principles, and this will be described in a future contribution.

Finally, besides the discovery of the seven new doubly-imaged QSOs and a binary QSO, the new selection principles leading to their discovery demonstrate the power of going beyond the survey catalogues alone and directly adopting images and more importantly, time-resolved data as expected from LSST and upcoming {\it Gaia} data releases, in searches for multiply-imaged, strongly lensed QSOs.

%%%%% ACKNOWLEDGEMENTS %%%%%
\begin{acknowledgements}
    AKM acknowledges the support from the Portuguese Funda\c c\~ao para a Ci\^encia e a Tecnologia (FCT) through grants SFRH/BPD/74697/2010, PTDC/FIS-AST/31546/2017, from the Portuguese Strategic Programme UID/FIS/00099/2013 for CENTRA, from the ESA contract AO/1-7836/14/NL/HB and from the Caltech Division of Physics, Mathematics and Astronomy for hosting research leaves during 2017-2018 and 2019, when the ideas and the  codes underlying this work were initially developed. MJG, SGD and AJD acknowledge a partial support from the NSF grants AST-1413600, AST-1518308, AST-1815034, and the NASA grant 16-ADAP16-0232. AMN acknowledges support from the NASA Postdoctoral Program Fellowship. The work of DS and AMN were carried out at the Jet Propulsion Laboratory, California Institute of Technology, under a contract with NASA. LD acknowledges support from the ESA PRODEX Programme `{\it Gaia}-DPAC QSOs' and from the Belgian Federal Science Policy Office. OW was supported by the Humboldt Research Fellowship for Postdoctoral Researchers. DD is supported by an Australian Government Research Training Program Scholarship. TM acknowledges the support of the Australian Research Council through grant FT150100099. We acknowledge partial support from `Actions sur projet INSU-PNGRAM', and from the Brazil-France exchange programmes Funda\c c\~ao de Amparo \`a Pesquisa do Estado de S\~ao Paulo (FAPESP) and Coordena\c c\~ao de Aperfei\c coamento de Pessoal de N\'ivel Superior (CAPES) -- Comit\'e Fran\c cais d'\'Evaluation de la Coop\'eration Universitaire et Scientifique avec le Br\'esil (COFECUB). This work has made use of the computing facilities of the Laboratory of Astroinformatics (IAG/USP, NAT/Unicsul), whose purchase was made possible by the Brazilian agency FAPESP (grant 2009/54006-4) and the INCT-A, and we thank the entire LAi team, specially Carlos Paladini, Ulisses Manzo Castello, Luis Ricardo Manrique and Alex Carciofi for the support. This project has received funding from the European Research Council (ERC) under the European Union’s Horizon 2020 research and innovation programme (grant agreement No 787886). We thank Rick White from Space Telescope Science Institute for the support with the PS1DR2 images. The Pan-STARRS1 Surveys (PS1) and the PS1 public science archive have been made possible through contributions by the Institute for Astronomy, the University of Hawaii, the Pan-STARRS Project Office, the Max-Planck Society and its participating institutes, the Max Planck Institute for Astronomy, Heidelberg and the Max Planck Institute for Extraterrestrial Physics, Garching, The Johns Hopkins University, Durham University, the University of Edinburgh, the Queen's University Belfast, the Harvard-Smithsonian Center for Astrophysics, the Las Cumbres Observatory Global Telescope Network Incorporated, the National Central University of Taiwan, the Space Telescope Science Institute, the National Aeronautics and Space Administration under Grant No. NNX08AR22G issued through the Planetary Science Division of the NASA Science Mission Directorate, the National Science Foundation Grant No. AST-1238877, the University of Maryland, Eotvos Lorand University (ELTE), the Los Alamos National Laboratory, and the Gordon and Betty Moore Foundation. The Dark Energy Camera Legacy Survey was conducted at the Blanco telescope, Cerro Tololo Inter-American Observatory, National Optical Astronomy Observatory (NOAO) (DECaLS; NOAO Proposal ID \# 2014B-0404; PIs: David Schlegel and Arjun Dey). The Legacy Surveys project is honored to be permitted to conduct astronomical research on Iolkam Du'ag (Kitt Peak), a mountain with particular significance to the Tohono O'odham Nation. The Australia Telescope Compact Array is part of the Australia Telescope National Facility which is funded by the Australian Government for operation as a National Facility managed by CSIRO. This work has made use of results from the ESA space mission {\it Gaia}, the data from which were processed by the {\it Gaia} Data Processing and Analysis Consortium (DPAC). Funding for the DPAC has been provided by national institutions, in particular the institutions participating in the {\it Gaia} Multilateral Agreement. The {\it Gaia} mission website is: http://www.cosmos.esa.int/gaia. Some of the authors are members of the {\it Gaia} Data Processing and Analysis Consortium (DPAC).
\end{acknowledgements}

\bibliographystyle{aa}
\bibliography{bibliography} 

\begin{appendix}
\label{sec:appendixSpectra}
\section{Spectra of the sources}
We present here the spectra of the multiply-imaged QSOs and of the likely double QSO discovered in this work.
 
\begin{figure*}[!htp]
\begin{center}
\includegraphics[width=1.8\columnwidth]{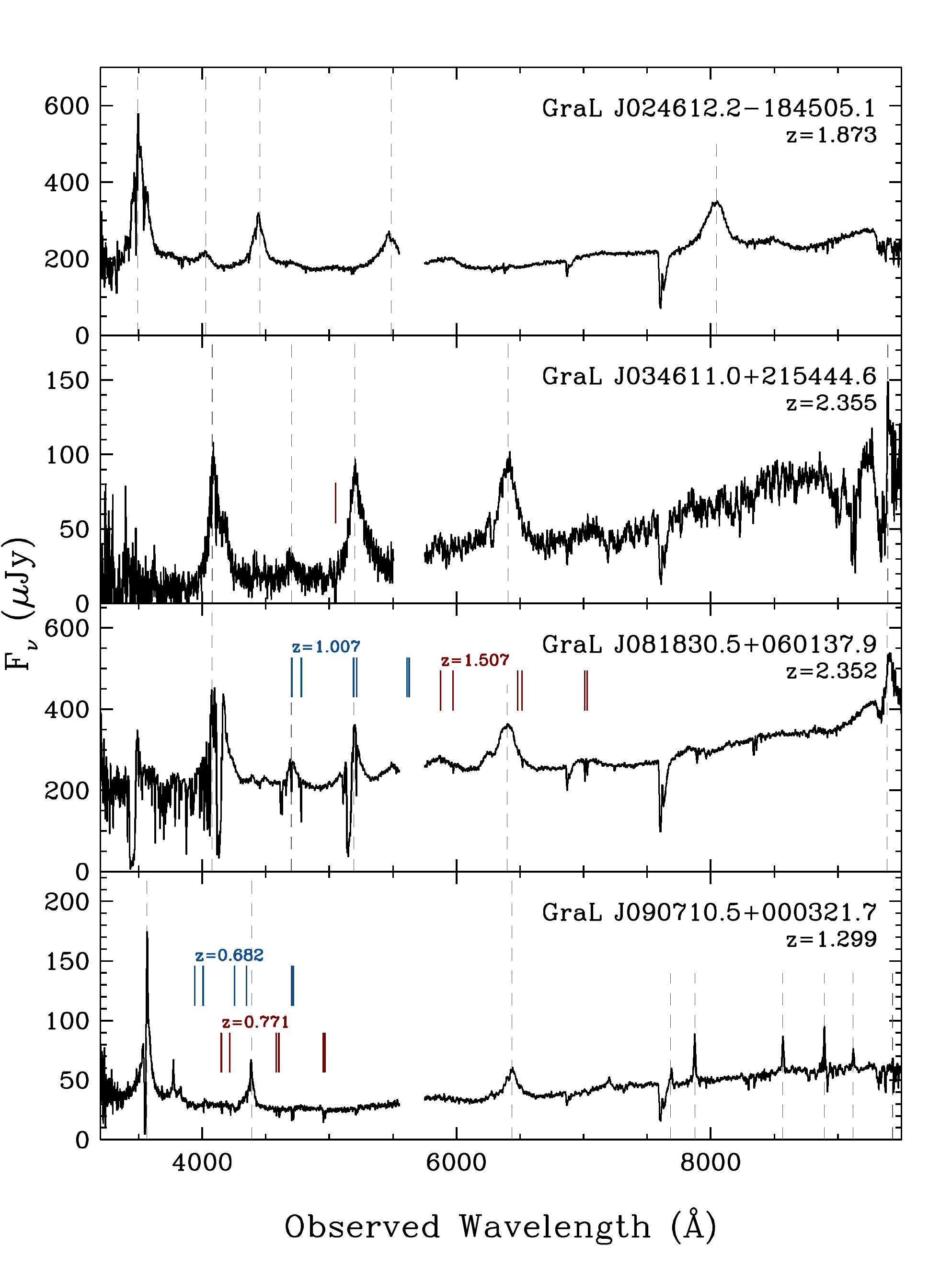}
\caption{Spectra of the first four sources presented herein.  Key emission lines are indicated with vertical dashed lines (i.e., in order, Ly$\alpha$, \sio, \civ, \ciii, \mgii, \neiii, \nev, \oii, \neiii, and H$\gamma$; given range of redshifts, only a subset of emission lines are detected for each source). Prominent absorption lines are presented with short, solid, vertical lines.  Labeled systems show absorption from \feii\ and \mgii.
\label{fig:spectra1}}
\end{center}
\end{figure*}

\begin{figure*}[!htp]
\begin{center}
\includegraphics[width=1.8\columnwidth]{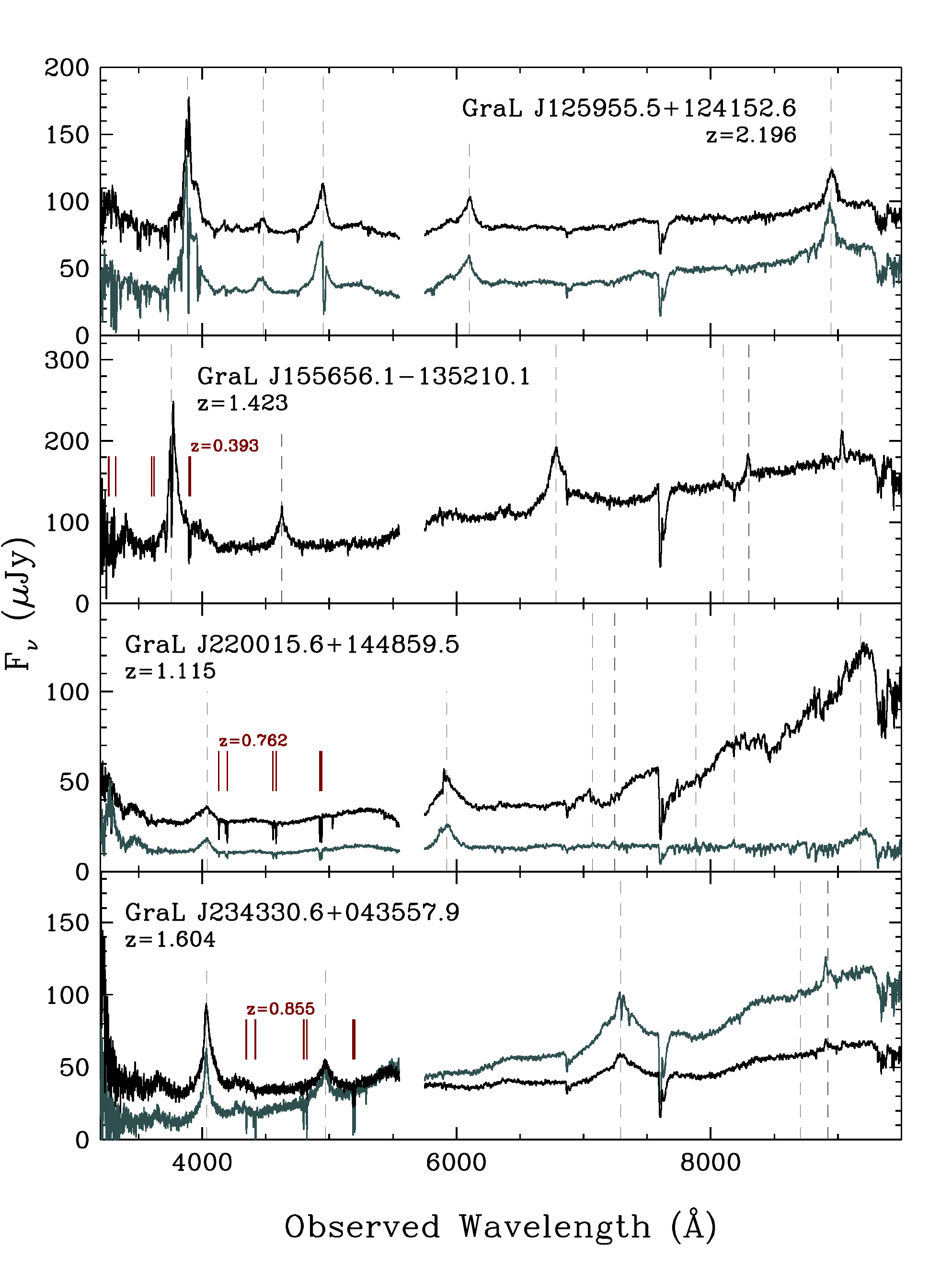}
\caption{Spectra of the second four sources presented herein.  Labels are as per Figure~A.1.  For sources where multiple components could be independently extracted, the two components are shown in different colors with an arbitrary additive offset included for visualization purposes.
\label{fig:spectra2}}
\end{center}
\end{figure*}

\end{appendix}
 
\end{document}